\documentclass[aps,prb,twocolumn]{revtex4}

\usepackage{graphicx,amsmath}

\begin{document}
\title{Superfluid density in the underdoped YBa$_2$Cu$_3$O$_{7-x}$:
Evidence for $d$-density wave order of pseudogap}
\author{M.R.~Trunin, Yu.A.~Nefyodov, A.F.~Shevchun}
\affiliation{Institute of Solid State Physics RAS, 142432 Chernogolovka,
Moscow district, Russia}

\begin{abstract}
The investigation of the penetration depth $\lambda_{ab}(T,p)$ in
YBa$_2$Cu$_3$O$_{7-x}$ crystals allowed to observe the following
features of the superfluid density $n_s(T,p)\propto
\lambda_{ab}^{-2}(T,p)$ as a function of temperature $T<T_c/2$ and
carrier concentration $0.078\le p\le 0.16$ in CuO$_2$ planes: (i)
$n_s(0,p)$ depends linearly on $p$, (ii) the derivative
$|dn_s(T,p)/dT|_{T\to 0}$ depends on $p$ slightly in the optimally
and moderately doped regions ($0.10<p\le 0.16$); however, it
rapidly increases with $p$ further lowering and (iii) the latter
finding is accompanied by the linear low-temperature dependence
$[-\Delta n_s(T)]\propto T$ changing to $[-\Delta n_s(T)]\propto
\sqrt{T}$. All these peculiarities can be treated in the framework
of $d$-density wave scenario of electronic processes in underdoped
high-T$_c$ materials.

\end{abstract}
\maketitle

Last years a lot of efforts were devoted to study the nature and
properties of pseudogap states of high-$T_c$ superconductors'
(HTSC) phase diagram. This area corresponds to lower concentration
$p$ of holes per copper atom and lower critical temperatures $T_c$
in comparison with the optimal value $p\approx 0.16$ and the
maximum temperature of the superconducting transition. In the
underdoped region HTSC strongly differ from conventional
materials, both in the normal and the superconducting states. This
difference is likely to occur in $p$- and $T$-dependences of the
superfluid density $n_s(T,p)$ of heavily underdoped HTSC.

It is well known that in clean BCS $d$-wave superconductors (DSC)
the dependence $\Delta n_s(T)\equiv n_s(T)-n_s(0)$ is linear on
temperature $T\ll T_c$: $[-\Delta n_s(T)]\propto T/\Delta_0$,
where $n_0\equiv n_s(0)$ and $\Delta_0\equiv \Delta (0)$ are the
superfluid density and the superconducting gap amplitude at $T=0$.
This dependence is confirmed by the measurements of the $ab$-plane
penetration depth $\lambda_{ab}(T)=\sqrt{m^*/\mu_0e^2n_s(T)}$:
$\Delta\lambda_{ab}(T)\propto T$ at $T<T_c/3$, where $\mu_0$,
$m^*$ and $e$ are the vacuum permeability, the effective mass and
the electronic charge, respectively. The derivative
$|\,dn_s(T)/dT|$ at $T\to 0$ determines $n_0/\Delta_0$ ratio. If
thermally excited fermionic quasiparticles are the only important
excitations even at $p<0.16$, then the slope of $n_s(T)$ curves at
$T\ll T_c$ is proportional to $n_0(p)/\Delta_0(p)$ ratio:
$|\,dn_s(T)/dT|_{T\to 0}\propto n_0(p)/\Delta_0(p)$. The
measurements of $\lambda_{ab}(0)$ in underdoped HTSC showed that
the superfluid density $n_0(p)\propto\lambda_{ab}^{-2}(0)$
increases approximately linearly with $p>0.08$ reaching its
maximum value at $p\approx 0.16$~\cite{Lor,Ber}.

When decreasing $p<0.16$ and hence approaching the dielectric phase, the
role of electron correlations and phase fluctuations becomes increasingly
significant. The generalized Fermi-liquid models (GFL) allow for this
through $p$-dependent Landau parameter $L(p)$~\cite{Lee,Mil1,Mil2} which
includes $n_0(p)$. The values of $\Delta_0(p)$ and $L(p)$ determine the
doping dependence of the derivative $|\,dn_s(T)/dT|_{T\to
0}=L(p)/\Delta_0(p)$. In Ref.~\cite{Lee} the ratio $L(p)/\Delta_0(p)$ does
not depend on $p$; the model~\cite{Mil1} predicts $L(p)/\Delta_0(p)\propto
p^{-2}$. The measurements of YBa$_2$Cu$_3$O$_{7-x}$ single
crystals~\cite{Bonn} and oriented powders~\cite{Pan} with the holes
concentration $p\gtrsim 0.1$ showed that the slope of $n_s(T)$ dependences
at $T\to 0$ is either slightly $p$-dependent~\cite{Bonn}, which agrees
with Ref.~\cite{Lee}, or diminishes with decreasing $p\leq
0.16$~\cite{Pan}, which contradicts the GFL models~\cite{Lee,Mil1,Mil2}.

Along with the above concept, there are a number of pseudogap
concepts~\cite{Timu,Tall,Sado,Norm} proposed in order to describe the
collapse of single-particle density of states near the Fermi level,
experimentally observed in underdoped HTSC at $T\gtrsim T_c$ by various
techniques. At microwave frequencies a breakdown of the normal skin-effect
condition in some HTSC was also treated in terms of pseudogap state
\cite{Sri1}. In Ref.~\cite{Ding} the pseudogap order parameter was found
to have the same $d$-wave symmetry as the superconducting one; it
influences the quasiparticles spectrum at $T<T_c$ as well. In the
precursor pairing model~\cite{Kost}, based on the formation of pair
electron excitations with finite momentum at $T>T_c$, this influence leads
to a rise of $\Delta_0(p)$ and decrease of $n_0(p)$ with $p$ lowering.
Hence, the decrease of the derivative $|\,dn_s(T)/dT|_{T\to 0}\propto
n_0(p)/\Delta_0(p)$ is expected. The $n_s(T,p)/n_0$ dependences calculated
in Ref.~\cite{Levi} show that their low temperature slopes decrease with
underdoping. An alternative behavior of $|\,dn_s(T)/dT|$ follows from
magnetic precursor $d$-density wave (DDW) scenario of pseudogap
\cite{Chak}. In this model a DDW order parameter $W(p,T)$ is directly
introduced into the quasi-particle band structure. At low energies the
excitation spectrum of DDW consists of conventional fermionic particles
and holes like that of DSC with which it competes at $p<0.2$. The DSC gap
$\Delta_0(p)$ steadily vanishes with $p$ decreasing, whereas the sum of
zero-temperature squares $\Delta_0^2(p)+W_0^2(p)$ remains constant
\cite{Tew}. In the issue, the DDW model predicts a growth of the slope of
the $n_s(T,p)/n_0$ curves at low $T$ and $p<0.1$.

The present paper aims at the experimental verification of the
theoretical predictions~\cite{Lee,Mil1,Mil2,Kost,Levi,Chak,Tew}.
To fulfil the task, we investigated $\lambda_{ab}(T)$ dependences
in YBa$_2$Cu$_3$O$_{7-x}$ single crystal with the oxygen
deficiency varied in the range $0.07\le x\le 0.47$. The
experiments were performed by the "hot-finger"
technique~\cite{Tru1} at the frequency of $\omega/2\pi=9.4$~GHz
and in the temperature range $5\le T\le 200$~K. The initial
YBa$_2$Cu$_3$O$_{6.93}$ crystal was grown in BaZrO$_3$
crucible~\cite{Erb} and had a rectangular shape with dimensions
being $1.6\times 0.4\times 0.1$~mm$^3$~\cite{Nef1}. To change the
carrier density, we successively annealed the crystal in the air
at various temperatures $T\ge 500^{\circ}$~C \cite{Tru2}. Finally,
five crystal states with critical temperatures $T_c=92, 80, 70,
57, 41$~K were investigated.
\begin{figure}[t]
\centerline{\includegraphics[width=0.9\columnwidth,clip]{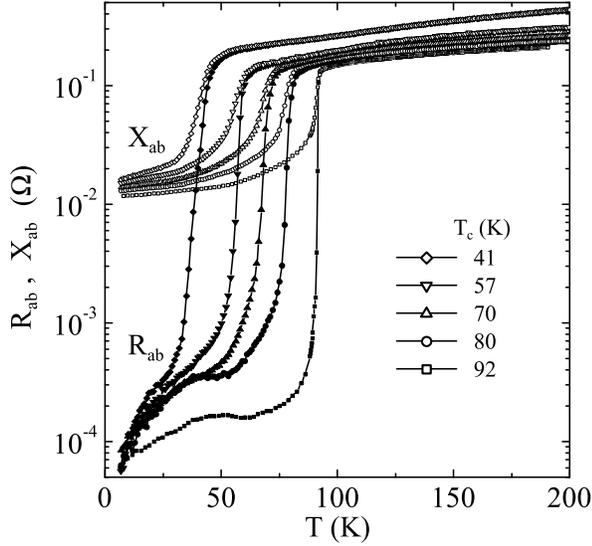}}
\caption{Real $R_{ab}(T)$ (solid symbols) and imaginary
$X_{ab}(T)$ (open symbols) parts of the $ab$-plane surface
impedance of YBa$_2$Cu$_3$O$_{7-x}$ single crystal for five
different $T_c$ values.} \label{f0}
\end{figure}
Using the empirical relation~\cite{Tal1}
$T_c=T_{c,max}[1-82.6(p-0.16)^2]$ with $T_{c,max}=92$~K at
$p=0.16$ ($x=0.07$), we get the concentrations $p=0.12, 0.106,
0.092, 0.078$ for other four states of YBa$_2$Cu$_3$O$_{7-x}$ with
lower $T_c$ values and $x=0.26, 0.33, 0.40, 0.47$ respectively.
According to $ac$-susceptibility measurements at the frequency of
100~kHz, superconducting transition width amounted to 0.1~K in the
optimally doped state ($p=0.16$); however, the width increased
with the decrease of $p$, having reached 4~K at $p=0.078$. The
temperature dependences of the $ab$-plane surface resistance
$R_{ab}(T)$ and reactance $X_{ab}(T)$ are shown in Fig.~1 for each
of the five crystal states. At $T<T_c/3$ all $R_{ab}(T)$ curves
are linear on $T$. The residual losses $R_{ab}(T\rightarrow0)$ do
not exceed 40~$\mu\Omega$. In more detail $R_{ab}(T)$ data will be
discussed elsewhere~\cite{Numssen}. The $X_{ab}(T)$ dependences in
Fig.~1 are constructed with allowance made for both (i) the
contribution $\Delta X_{ab}^{th}(T)$ of thermal expansion of the
crystal which essentially affects the measured reactance shift
$\Delta X_{ab}(T)$ at $T>0.9\,T_c$, and (ii) the additive constant
$X_0$ which is equal to the difference between the values of
$[\Delta X_{ab}(T)+\Delta X_{ab}^{th}(T)]$ and $R_{ab}(T)$ at
$T>T_c$: $X_{ab}(T)=\Delta X_{ab}(T)+\Delta
X_{ab}^{th}(T)+X_0$~\cite{Nef1}. So, in the normal state for each
of the five crystal states we have $R_{ab}(T)=X_{ab}(T)$ which
implies the validity of the normal skin-effect condition. This
finding enables to extract the absolute values of the $ab$-plane
penetration depth $\lambda_{ab}(T)=X_{ab}(T)/\omega\mu_0$ from
$X_{ab}(T)$ curves at $T<T_c$.

Fig.~2 shows the low temperature sections of $\lambda_{ab}(T)$
curves. The linear extrapolation (dashed lines) of these
dependences at $T<T_c/3$ gives the following $\lambda_{ab}(0)$
values: 152, 170, 178, 190, 198~nm for $p=0.16, 0.12, 0.106,
0.092, 0.078$, respectively. The error in $\lambda_{ab}(0)$ is
largely determined by the measurement accuracy of the additive
constant $X_0$. In our experiments the root-mean-square difference
between $R_{ab}(T)$ and $X_{ab}(T)$ in the normal state
corresponded to about 5~nm accuracy in $\lambda_{ab}(0)$ value.
\begin{figure}[t]
\centerline{\includegraphics[width=0.9\columnwidth,clip]{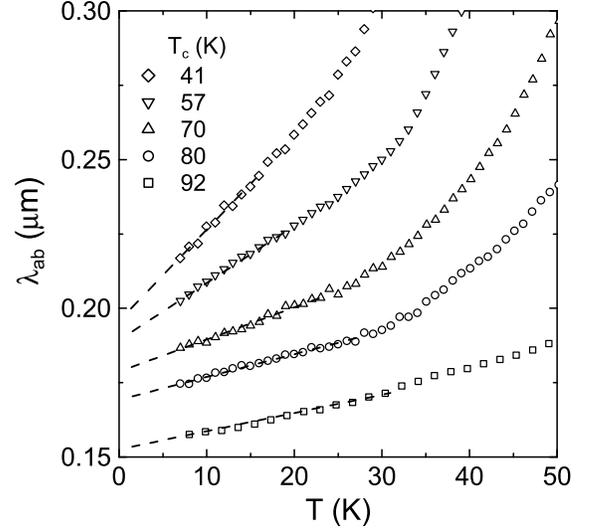}}
\caption{Low-temperature dependences of $\lambda_{ab}(T)$ (open
symbols) measured for five states of YBa$_2$Cu$_3$O$_{7-x}$
crystal with $T_c=92$~K, $T_c=80$~K, $T_c=70$~K, $T_c=57$~K, and
$T_c=41$~K. Dashed lines are linear extrapolations at $T<T_c/3$.}
\label{f1}
\end{figure}

\begin{figure}[b]
\centerline{\includegraphics[width=0.9\columnwidth,clip]{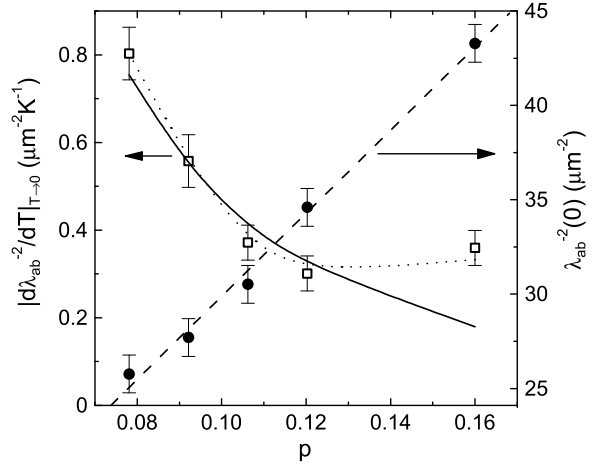}}
\caption{The values of $\lambda_{ab}^{-2}(0)=n_s(0)\mu_0e^2/m^*$
(right scale) and slopes $|\,d\lambda_{ab}^{-2}(T)/dT|_{T\to
0}=\mu_0e^2/m^*|\,dn_s(T)/dT|_{T\to 0}$ (left scale) as a function
of doping $p=0.16-\sqrt{(1-T_c/T_{c,max})/82.6}$ with
$T_{c,max}=92$~K in YBa$_2$Cu$_3$O$_{7-x}$. Error bars correspond
to experimental accuracy. The dashed and dotted lines guide the
eye. The solid line is $|\,dn_s(T)/dT|\propto p^{-2}$ dependence.}
\label{f2}
\end{figure}

As follows from Fig.~3, halving of the concentration (namely, from
$p=0.16$ to $p=0.078$) results in approximately two times smaller
$\lambda_{ab}^{-2}(0)=n_0\mu_0e^2/m^*$ value. Similar behavior
$n_0(p)\propto p$ within the range $0.08<p\le 0.16$ was observed
by other groups~\cite{Lor,Ber}. It is easily seen that this
dependence contradicts Uemura's relation $n_0(p)\propto T_c(p)$
\cite{Uem}. The naive linear extrapolation of the dashed line in
Fig.~3 at $p<0.08$ leads to nonphysical result: $n_0(p)$ is finite
at vanishing $p$. To the best of our knowledge there is no data of
superfluid density measurements in HTSC at $p<0.08$. As for
theoretical predictions, $n_0$ linearity on $p$ extends down to
$p=0$ in the model~\cite{Lee}, while in the DDW
scenario~\cite{Tew,Wan} it exists in the underdoped range of phase
diagram where the DSC order parameter grows from zero to its
maximal value (Fig.~1 from Ref.~\cite{Wan}). The latter agrees
with our data.

In Fig.~3 we also show the slopes
$|\,d\lambda_{ab}^{-2}(T)/dT|_{T\to 0}\propto |\,dn_s(T)/dT|_{T\to
0}$ of $\lambda_{ab}^{-2}(T)$ curves obtained from
$\lambda_{ab}(T)$ dependences at $T<T_c/3$. The value of
$|\,d\lambda_{ab}^{-2}(T)/dT|$ changes slightly at $0.1<p\leq
0.16$ in accordance with~Ref.\cite{Lee}. However, it grows
drastically at $p\lesssim 0.1$, namely, $\lambda_{ab}^{-2}(T)$
slope increases 2.5 times with $p$ decrease from 0.12 to 0.08.
$|\,d\lambda_{ab}^{-2}(T)/dT|\propto p^{-2}$
dependence~\cite{Mil1} is shown by solid line in Fig.~3 and
roughly fits the data at $p\leq 0.12$. The dotted line drawn
through $|\,d\lambda_{ab}^{-2}(T)/dT|$ experimental points in
Fig.~3 qualitatively agrees with the behavior of this quantity in
the DDW model~\cite{Tew,Wan}.
\begin{figure}[t]
\centerline{\includegraphics[width=0.9\columnwidth,clip]{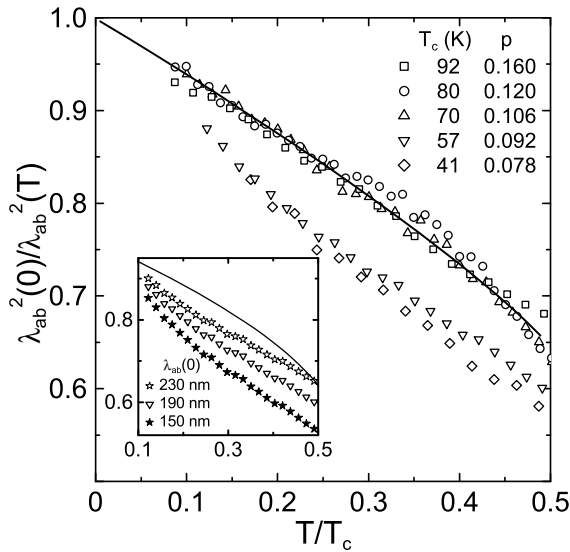}}
\caption{The measured dependences of
$\lambda_{ab}^2(0)/\lambda_{ab}^2(T)=n_s(T)/n_s(0)$ at $T<T_c/2$
in YBa$_2$Cu$_3$O$_{7-x}$ with different doping. The solid line is
the $\lambda_{ab}^2(0)/\lambda_{ab}^2(T)$ dependence in BCS
$d$-wave superconductor (DSC). The inset shows $n_s(T)/n_0$
experimental curve for $p=0.092$ and the ones obtained using
$\lambda_{ab}(0)$ increased (open stars) and decreased (solid
stars) by 40~nm.} \label{f3}
\end{figure}

The temperature dependence of the superfluid density $n_s(T)$ at
low $T$ in the heavily underdoped YBa$_2$Cu$_3$O$_{7-x}$ proves to
be one more check-up of the DDW-scenario of pseudogap.
$\lambda_{ab}^2(0)/\lambda_{ab}^2(T)=n_s(T)/n_0$ dependences
obtained from the data in Fig.~2 are shown in Fig.~4 for different
values of $p$. The solid line represents the DSC result. The
evident peculiarities in Fig.~4 are the concavity of $n_s(T)/n_0$
curves corresponding to the heavily underdoped states ($p=0.078$
and $p=0.092$) and their deviation from DSC and the curves for
$p=0.16, 0.12, 0.106$. It should be noted that these peculiarities
do not strongly depend on $\lambda_{ab}(0)$ values. This is
demonstrated in the inset to Fig.~4, where $n_s(T)/n_0$
experimental curve for $p=0.092$ is compared to the ones obtained
using $\lambda_{ab}(0)$ increased (open stars) and decreased
(solid stars) by 40~nm. Actually, the latter value is much higher
than the experimental uncertainty.
\begin{figure}[t]
\centerline{\includegraphics[width=0.99\columnwidth,clip]{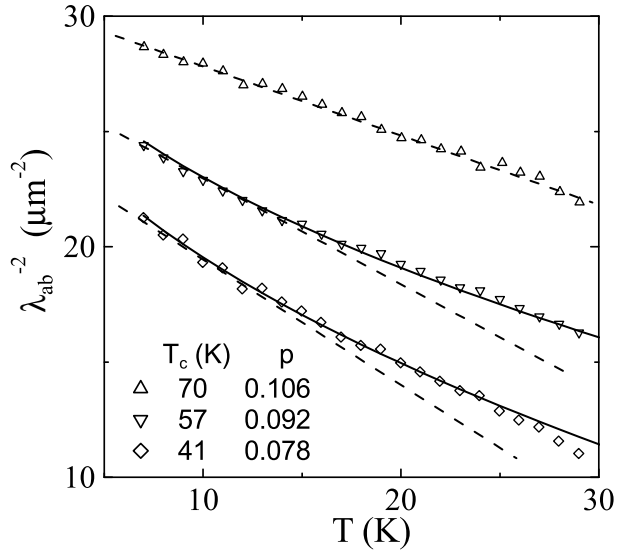}}
\caption{Comparison of experimental $\lambda_{ab}^{-2}(T)\propto
n_s(T)$ curves (symbols) with linear
$[-\Delta\lambda_{ab}^{-2}(T)]\propto T$ (dashed lines) and root
$[-\Delta\lambda_{ab}^{-2}(T)]\propto \sqrt T$ (solid lines)
dependences for moderately doped ($p=0.106$, $x=0.33$) and heavily
underdoped ($p=0.092$, $x=0.40$; $p=0.078$, $x=0.47$)
YBa$_2$Cu$_3$O$_{7-x}$.} \label{f4}
\end{figure}

The behavior of the superfluid density $n_s(T)/n_0$ in Fig.~4
contradicts the conclusions of the precursor pairing
model~\cite{Levi}, but agrees with the DDW scenario~\cite{Tew}.
According to Ref.~\cite{Tew}, at temperatures much smaller than
the relevant energy scales $W_0$ and $\Delta_0$, only the nodal
regions close to the points ($\pi/2,\pi/2$) and symmetry-related
points on the Fermi surface will contribute to the suppression of
the superfluid density. In a wide range of temperatures the
$n_s(T)$ dependence will be linear for the optimally and
moderately doped samples, in which $\Delta_0$ is larger than or
comparable to $W_0$ and plays a leading role in the temperature
dependence of the superfluid density. However, for the heavily
underdoped samples the situation is quite different. Though in the
asymptotically low-temperature regime the suppression of the
superfluid density is linear on temperature, there is an
intermediate temperature range over which the suppression actually
behaves as $\sqrt T$. It is worth emphasizing that the authors of
Ref.~\cite{Tew} state that these features are independent of the
precise $W_0(p)$ and $\Delta_0(p)$ functional forms. The only
input that is needed is the existence of DDW order which
diminishes with $p$ increase and complementary development of the
DSC order. The DDW order eats away part of the superfluid density
from an otherwise pure DSC system. Actually, in the intermediate
temperature range $0.1\,T_c<T\lesssim 0.5\,T_c$ the experimental
$n_s(T)$ curves in YBa$_2$Cu$_3$O$_{6.60}$ and
YBa$_2$Cu$_3$O$_{6.53}$ with $p<0.1$ are not linear but similar to
$\sqrt{T}$-dependences. This is confirmed by Fig.~5, where the
measured curves $\lambda_{ab}^{-2}(T)\propto n_s(T)$ are compared
with linear ($\propto T$) in YBa$_2$Cu$_3$O$_{6.67}$ ($p=0.106$)
and $\sqrt{T}$-dependences $\Delta\lambda_{ab}^{-2}(T)=-3\sqrt{T}$
($\lambda_{ab}$ and $T$ are expressed in $\mu$m and K) in
YBa$_2$Cu$_3$O$_{6.60}$ ($p=0.092$) and
$\Delta\lambda_{ab}^{-2}(T)=-3.5\sqrt{T}$ in
YBa$_2$Cu$_3$O$_{6.53}$ ($p=0.078$). Dashed lines in Fig.~5
correspond to the linear at $T<T_c/3$ dependences of
$\lambda_{ab}(T)$ presented in Fig.~2 and extended to higher
temperatures. It is also interesting that these peculiarities of
$\Delta\lambda_{ab}^{-2}(T)$ dependences in
YBa$_2$Cu$_3$O$_{6.60}$ and YBa$_2$Cu$_3$O$_{6.53}$ are
accompanied by inflection of the resistivity $\rho_{ab}(T)$ curves
in the normal state of these samples which can be illustrated by
Fig.~2 from Ref.~\cite{Tru2}.

Thus, three main experimental observations of this paper, viz, (i)
linear dependence of $n_0(p)$ in the range $0.078\le p\le 0.16$,
(ii) drastic increase of low-temperature $n_s(T)$ slope at
$p<0.1$, and (iii) the deviation of $\Delta n_s(T)$ dependence
from universal BCS behavior $[-\Delta n_s(T)]\propto T$ at
$T<T_c/2$ towards $[-\Delta n_s(T)]\propto \sqrt{T}$ with
decreasing $p<0.1$, evidence the DDW scenario~\cite{Chak,Tew,Wan}
of electronic processes in underdoped HTSC. Nevertheless, the
measurements of $n_s(T)$ at lower temperatures and in the
high-quality samples with smaller carrier density are necessary
for ultimate conclusion.

Helpful discussions with A.I.~Larkin and Sudip Chakravarty are gratefully
acknowledged. This research was supported by RFBR grants Nos. 03-02-16812
and 02-02-08004.

\end{document}